\documentclass[aps,prl,twocolumn,showpacs,superscriptaddress,groupedaddress]{revtex4}  
\usepackage{graphicx}  
\usepackage{dcolumn}   
\usepackage{bm}        
\usepackage{amssymb}   
\usepackage{cases}
\usepackage{subfigure}
\begin{document}

\widetext

  \centerline{\em D\O\ INTERNAL DOCUMENT -- NOT FOR PUBLIC
DISTRIBUTION}


\title{Validity of Transport Energy in Disordered Organic Semiconductors}
\author{Ling Li, Nianduan Lu, and Ming Liu}
\email{lingli@ime.ac.cn, liuming@ime.ac.cn}
\altaffiliation{Institute of microelectronics, Chinese Academy of
Sciences, Beijing, 100029, China}

\date{\today}

\begin{abstract}
A systematic study of the transport energy in disordered organic
semiconductors based on variable range hopping theory has been
presented here. The temperature, electric field, material disorder
and carrier concentration dependent transport energy is extensively
discussed. We demonstrate here, transport energy is not a general
concept and invalid even in low electric field and concentration
regime.
\end{abstract}

\maketitle


Understanding the charge transport mechanism in disordered organic
semiconductors such as conjugated and molecularly doped polymers, is
of crucial importance to design and synthesize better materials.
Physical transport phenomena in organic semiconductors are in
general very complex and arise from energy and spatial disorder,
which invalidates the use of the traditional band transport theories
based on the periodic distribution of atoms in inorganic crystals.
Currently, charge transport in organic semiconductors is often
described in terms of  hopping theory \cite{mott}. A fundamental
work in the modeling of the hopping transport in organic
semiconductors is the Gaussian disorder model proposed by Bassler
{\cite{bassler1,bassler2}}. Subsequent to this Monte Carlo
simulation, a number of theoretical investigations concerning charge
transport in amorphous organic systems utilizing the Gaussian
approach have been published and considerable progress in the
analytical description of the problem has been made
{\cite{novikov,dunlap,arkhipov1,li1}}. An important simplification
of the complex hopping transport mechanism was the introduction of
the transport energy concept, This concept allows the complex
hopping mechanism in the band tails to be interpreted in terms of a
multiple-trapping-and-release model, where the transport energy
plays the role similar to the mobility edge in amorphous inorganic
semiconductors
{\cite{monro,arkhipov2,li2,baranovskii1,baranovskii2,roland1}}.
Thermally stimulated luminescence{\cite{bassler3}}, carrier
dependent mobility{\cite{blom1,baranovskii3,li2,vissenberg}},
seebeck coefficient \cite{germs}, as well as injection
phenomena{\cite{arkhipov3}}, have already been described utilizing
this transport energy concept. Despite the concept of transport
energy has widely been applied to describe the charge hopping
transport of different organic semiconductors, the validity of this
concept is nontrivial. As transport energy is derived on the basis
of zero electric field and low carrier concentration regime, and
only the hopping upwards transport is concerned. However, in the
real disordered organic semiconductors devices, the electric field
could reach $1\times 10^8 V/m$  and carrier concentration could as
high as $1\times 10^{26}m^{-3}$,therefore, the applicability of the
transport energy to real disordered organic semiconductors should be
a matter of intensive research, where it is important to distinguish
whether the organic semiconductor exist the mobility edge (transport
energy) or whether this concept could be used to describe the
mobility in hopping system. In this letter, we show how the
transport energy changes with temperature, electric field, carrier
concentration and material disorder. Using this analysis, we
conclude that the transport energy is not a general concept and
invalid in real disordered organic semiconductors.

\textit{Model}.---Generally speaking, there are two ways to define
he concept of transport energy in organic semiconductors, both of
them are based on the Miller-Abrahams expressions \cite{miller}, the
hopping transport takes place via tunneling between an initial
states $i$and a target states$j$.The tunneling process is described
as

\begin{equation}
  \nu=\nu_0\exp\left(-u\right)=\nu_0\left\{
   \begin{array}{c}
   \exp\left(-2\alpha
R_{ij}-\frac{E_j-E_i}{kT}\right),  E_i>E_j\\
  \left(-2\alpha R_{ij}\right).  \qquad\qquad\qquad E_i<E_j\\
   \end{array}
  \right.
  \end{equation}
Here, $\nu_0$ is the attempt-to-jump frequenct, $R_{ij}$ is the
hopping distance, $u$ is the hopping range \cite{arkhipov2},$E_i$
and $E_j$ are the eneries at site $i$ and $j$, respectively,
$\alpha$ is the inverse localized length and $k$ is the Boltzmann
constant. Then, if only the hopping upwards $\nu_{\uparrow}$ is
taken into account, the transport energy is defined as the finial
energy $E_t$ that the hopping transport has maximum rate, so $E_t$
could be obtained by the equation as \cite{baranovskii2}
\begin{equation}
\frac{\partial\nu_{\uparrow}}{\partial E_t}=0.
\end{equation}
This gives
\begin{equation}
g\left(E_t\right)\left[\int^{E_t}_{-\infty}g\left(E\right)dE\right]^{-4/3}=\frac{1}{\alpha
kT}\left(\frac{9\pi}{2}\right)^{1/3}.
\end{equation}
Where $g\left(E\right)$ is the density of states (DOS). Here we can
see that the hopping transport is limited by upward transitions from
filled states to empty states and the transport energy has been
defined as the preferred energy to which the fastest upward
transitions occur.\\
In the other case, the transport energy is derived based on the
average number of target sites $n\left(E_i,u\right)$ as
\begin{equation}
n\left(E_i,u\right)=\int_0^{u/{2\alpha}}R_{ij}^2dR_{ij}\int_{-\infty}^{E_i+kT\left(u-2\alpha\right)}g\left(E\right)dE.
\end{equation}
If we disregard the hopping downwards and choose the starting energy
$E_i$ as $-\infty$, the transport energy could be obtained by
setting $n\left(E_i,u\right)=1$ and the result is \cite{arkhipov2}
\begin{equation}
\int_{-\infty}^{E_t}g\left(E\right)\left(E_t-E\right)^3dE=\frac{6}{\pi}\left(\alpha
kT\right)^3.
\end{equation}
This definition hints that the transport energy is the site energy
to which the hopping upwards need the least energy in energy
space.\\
When there exists an electric field $F$, the electric field will
lower the Coulomb barrier, which leads to the reduction of the
thermal activation energies, and the hopping range with normalized
energy can therefore be rewritten as \cite{apsley,li3}
\begin{equation}
 u=\left\{
   \begin{array}{c}
 2\alpha\left(1+\beta\right)
R_{ij}+\epsilon_j-\epsilon_i,  \epsilon_j>\epsilon_i-\beta\cos\theta\\
  \left(2\alpha R_{ij}\right).  \qquad\qquad\qquad \epsilon_j<\epsilon_i-\beta\cos\theta\\
   \end{array}
  \right.
\end{equation}
Where $\beta=Fe/{2\alpha kT}$ and and $\theta$ is the angle between
$R_{ij}$ and the electric field ranging from $0$ to $pi$.For a site
with energy $\epsilon_i$ in the hopping space, the most probable hop
for a carrier on this site is to an empty site at a range $u$, where
it needs the minimum energy. Conduction is the result of a long
sequence of hops through this hopping space. Then, following the
method in \cite{apsley}, we derive the number of empty sites
enclosed by the constant range $u$, as
\begin{eqnarray}
n\left(\epsilon_i,u,F\right)
=\nonumber\qquad\qquad\qquad\qquad\qquad\qquad\qquad\qquad\qquad\\
\frac{1}{8\alpha^3}\int_0^{\Pi}d\theta\sin\theta\int_0^udr2\pi
r^2\int_{-\infty}^{u+\epsilon_i-r\left(1+\beta\cos\theta\right)}d\epsilon
\nonumber\\\times
g\left(\epsilon\right)\left[1-f\left(\epsilon,\epsilon_F\right)\right].
\end{eqnarray}
Here $f\left(\epsilon,\epsilon_F\right)$ is the Fermi Dirac
distribution and $1-f\left(\epsilon,\epsilon_F\right)$ is the
probability that the finial site is empty, the Fermi energy
$\epsilon_F$ is calculated by the condition
\begin{equation}
c=\int_{-\infty}^{\infty}d\epsilon\frac{
g\left(\epsilon\right)}{1+exp\left(\epsilon-\epsilon_F\right)}.
\end{equation}
Here $c$ is the carrier concentration. By change the integration
variable, equation (7) will be in the form of
\begin{eqnarray}
n\left(\epsilon, u,
F\right)=\frac{2\pi}{3\times8\alpha^3}\int_{-1}^{1}d\tau\qquad\qquad\qquad\qquad\nonumber\\
\times\left[\int_{\epsilon_i-\beta u\tau}^{\epsilon_i+u}d\epsilon
g\left(\epsilon\right)\frac{\left(u+\epsilon_i-\epsilon\right)^3}{\left(1+\beta\tau\right)^3}+\int_{-\infty}^{\epsilon_i-\beta
u\tau}d\epsilon g\left(\epsilon\right)\right]
\end{eqnarray}

The first term on the right-hand side of equation (9) gives the
number of shallower states,and the second one describes the number
of target states which are deeper than the starting site . According
to the variable range hopping theory
\cite{apsley,arkhipov1,baranovskii3}, at a given field and
temperature, almost every starting localized state has only one well
distinguished nearest empty target hopping neighbor, i.e. another
localized state that is characterized by the minimum value of the
hopping range, this range could be obtained by solving the equation
\begin{equation}
n\left(\epsilon_i, F, u\right)=1.
\end{equation}
This equation established the basis of our model. the target site
energy for every hopping process can be well evaluated from the
euqation (10).

\textit{Validity of Arkhipov transport energy.}---To check this
model validity, we disregard the downwards hopping as well, equation
(9) then reads as
\begin{eqnarray}
1=\frac{2\pi}{3\times8\alpha^3}\int_{-1}^{1}d\tau\int_{\epsilon_i-\beta
u\tau}^{\epsilon_i+u}d\epsilon
g\left(\epsilon\right)\frac{\left(u+\epsilon_i-\epsilon\right)^3}{\left(1+\beta\tau\right)^3}
\end{eqnarray}
 Using the developed model equation (11),
 we now proceed to calculate the relation between starting energy
 and finial energy in real disordered organic semiconductors. We
 take the Gaussian form
 $g\left(\epsilon\right)=\frac{N_t}{\sqrt{2\pi}\sigma}\exp\left(-\frac{\epsilon^2}{2\sigma^2}\right)$
 density of states in the full manuscript, where $N_t$ is the number of states per unit volume and $\sigma=\sigma_0/kT$ indicates the width of the
 DOS. $N_t=1\times 10^{28}m^{-3}$ is taken in the full manuscript, a typical value for the
 relevant organic semiconductors. It is instructive to calculate this target energy as a function
of the initial energy for different carrier concentration,
corresponding to different fermi-level, the results are displayed in
the insert of Fig. 1. A clear observation is that, in this
situation, the transport energy does exist for
 deeper starting energy but increases with the carrier
 concentration. Field dependent transport energy is plotted in the
  Fig.1, the same as concentration dependent transport energy, field
 does not change transport energy for deeper energies but decreases
 for field higher than $1\times 10^{7}V/m$, this field strength is
 actually low field for most organic devices. The reason for the
 decrease of transport energy is that, the electric field can change the
 energy difference between the finial and target sites, and thereby,
 assist carrier jumps along the field direction, so the target energy will decrease
 on the contrary. For the very low
 carrier concentration, the field and concentration dependent transport energy could
 be derived as
 \begin{eqnarray}
n\left(\epsilon, u, F\right)
\approx\frac{2\pi}{24\alpha^3\left(1-\beta^2\right)^2}\int_{\epsilon_i-\beta
u}^{\epsilon_i+u}g\left(\epsilon\right)\left(\epsilon_i+u-\epsilon\right)^3d\epsilon\nonumber\\
=\frac{2\pi}{24\alpha^3\left(1-\beta^2\right)^2}\int_{-\infty}^{\epsilon_t}g\left(\epsilon\right)\left(\epsilon_t-\epsilon\right)^3d\epsilon\qquad\qquad
\end{eqnarray}
Certainly, this is not a general concept as well.\\
We then consider the effect of the lattice disorder, which is caused
by the random molecular
  packing in organic semiconductors \cite{hulea}. The relation
 between transport energy and materials disorder $\sigma$ is shown
 in Fig. 2 (a). An important result is that, the transport energy is
 invalid for higher energies, but these energies make sense for the
 charge transport in real organic semiconductors. It is well known,
 in the low carrier concentration
 regime, the hopping usually jumps from the so-called equilibrium energy $E_{\infty}=-\left(\sigma\right/kT)^2=-25kT$ for
 $\sigma/kT=5$; In the high carrier concentration regime, the
 carrier jumps from Fermi energy ($\epsilon_F=-20k_BT$ here). An obvious feature appears here is the transport energy
 does not exist for energy higher than $-18kT$ in the case of $\sigma/kT=5$. In organic semiconductors, $\sigma=5kT$
 correspond to $\sigma=0.12ev$ at room temperature, a typical value for organic semiconductors \cite{blom3}. Therefore, the transport energy
 has no generality and is invalid for real organic semiconductor
 system. Please note that the field used here is $1\times 10^5 V/m$,
  which is low field regime; the fermi energy chosen here is also corresponding
 to the low concentration regime. The other
 parameters values are also typical ones for organic semiconductors
 such as $\alpha=1nm$. To investigate the reason for this phenomena,
 we plot the ratio between downwards and upwards hopping in Fig. 2(b),
 one can see clearly that, the downwards hopping has little effect on
 the charge transport characteristics and should not account for this
 result. One possible explanation is, for the
 small disorder Gsussian DOS, since the carriers have few nearest
 neighbor sites to choose and it is reasonable to hop to the same
 energy; But for the wider Gaussian DOS, carriers has more near
 neighbor sites and more active, hence the target energy will be
 random according to  the calculation.
 Therefore, the approximation in \cite{baranovskii2,arkhipov1} that the
 the  transport energy in zero field is independent the starting
 energy is not correct in real organic semiconductors.


\begin{figure}[h]
             \centering \scalebox{0.28}{\includegraphics{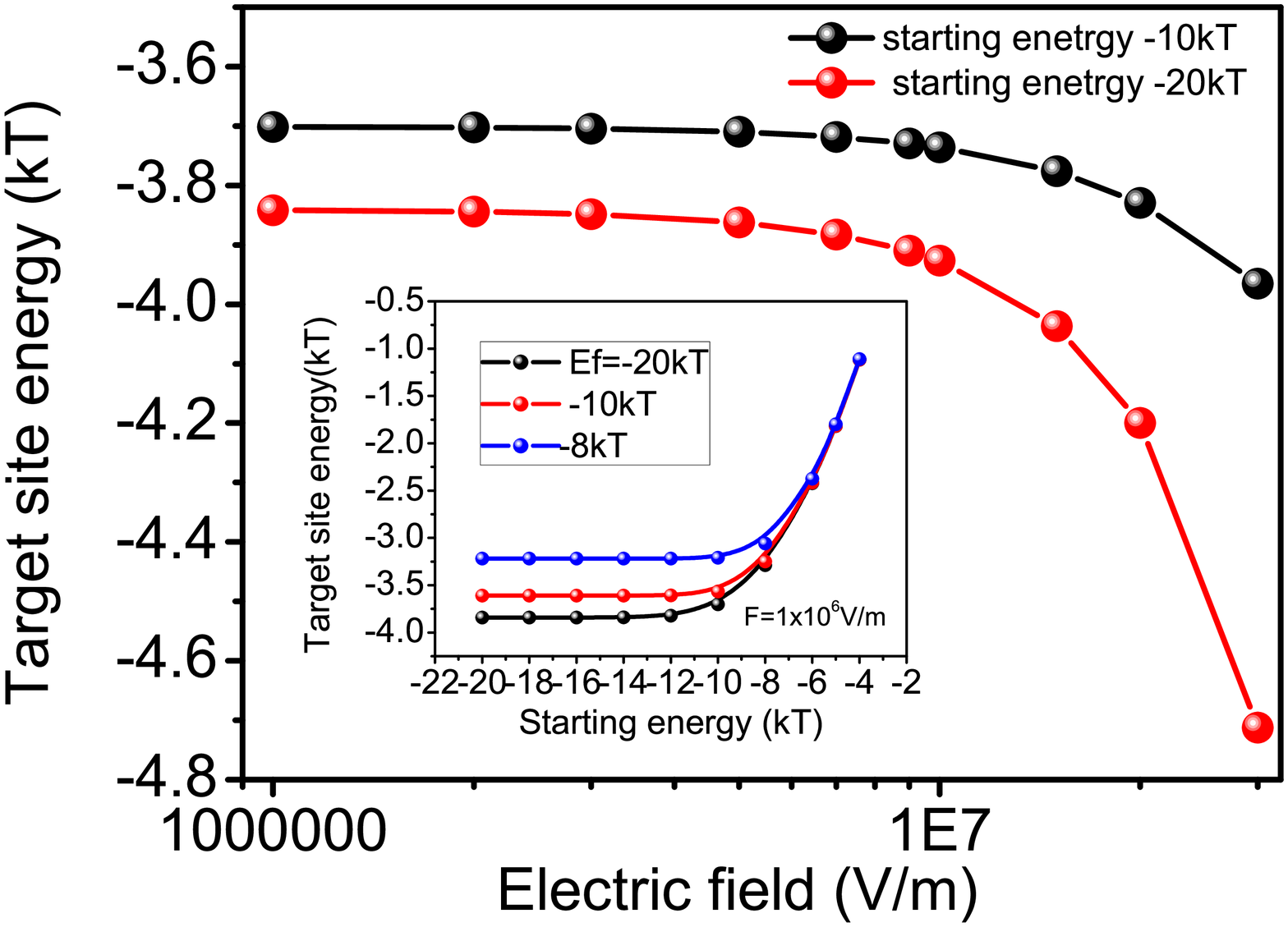}}
             \vspace*{-7pt}\caption{The computed transport energy as function of the
             starting energy
for different electric field . The inset shows the carrier
concentration (Fermi level) dependent transport energy.}
             \end{figure}

\begin{figure}
\centering \subfigure[]{
\label{fig:subfig:a} 
\includegraphics[width=2.6in]{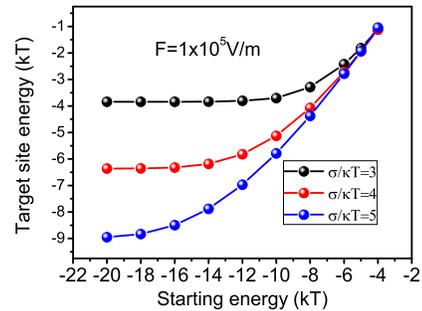}}
\hspace{1in} \subfigure[]{
\label{fig:subfig:b} 
\includegraphics[width=2.8in]{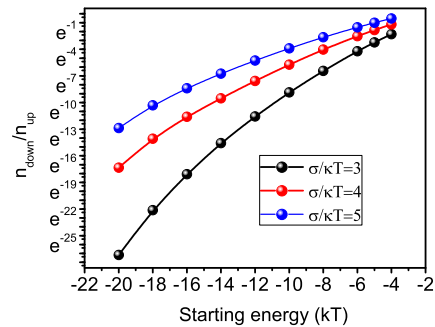}}
\caption{(a) The disorder dependent transport energy for different
starting energy. (b) The ratio of hopping-downwards site number and
hopping upwards number}
\label{fig:subfig} 
\end{figure}
The question arises now on how much difference between the
calculated mobility using Akhipov method and the work here. The
comparison is shown in Fig. 3. It is clearly seen that, in the low
temperature regime, Akhipov model will underestimate the mobility
but overestimate the mobility in high temperature. And this trend is
even more pronounced  with carrier concentration increasing.

\textit{Validity of Baranovskii transport energy model.}---According
to the definition of Baranovskii transport energy, the hopping
upwards rate has maximum value for every jumping, hence, we
derivative equation (6) as
\begin{equation}
\left \{
\begin{array}{rl}
    \frac{\partial\nu_\uparrow}{\partial\epsilon_j}=2\alpha\left(1+\beta\right)\frac{\partial R_{ij}}{\partial\epsilon_j}+1=0, \\
    \frac{\partial\nu_\uparrow}{\partial\theta}=2\alpha\left(1+\beta\right)\frac{\partial R_{ij}}{\partial\theta}=0,  \\
    1=4\pi R_{ij}^2\int_{0}^{\theta}\sin\theta d\theta\int_{-\infty}^{\epsilon_j}d\epsilon
g\left(\epsilon\right)\left[1-f\left(\epsilon,\epsilon_F\right)\right].\\
\end{array}
\right.
\end{equation}
The $R_{ij}$ that is the function of $\theta$ and $\epsilon_j$,
could be obtained as the equation (11). Connecting Gaussian DOS and
equation(12), we numerically calculate the transport energy for this
definition. The concentration dependent transport energy is shown in
Fig. 4. The deviation of target energy from the transport energy is
more dramatic, even at the starting energy $\epsilon_i=24k_BT$, the
transport energy is obvious invalid. This holds for the field
dependent transport energy (the insert figure), too. We also find,
the same conclusion is obtained by introducing the percolation
parameter \cite{baranovskii2} in equation (13) .

             \begin{figure}[h]
             \centering \scalebox{0.53}{\includegraphics{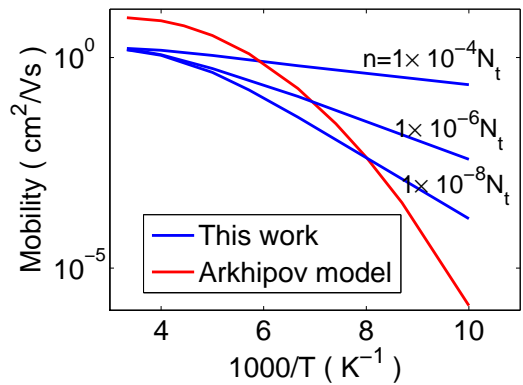}}
             \vspace*{-7pt}\caption{The Comparison between the calculated
             mobility in this work ($F=5\times 10^4 V/cm$) and Akhipov mobility model for different temperatures.}
             \end{figure}

             \begin{figure}[h]
             \centering \scalebox{0.28}{\includegraphics{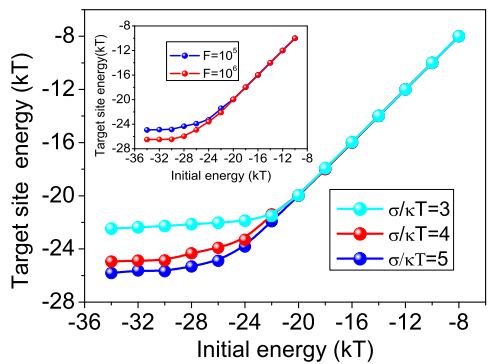}}
             \vspace*{-7pt}\caption{The computed Baranovskii transport energy as function of the
             starting energy
for different electric field . The inset shows the electric field
dependent Baranovskii transport energy.}
             \end{figure}
 In conclusion, the validity of the
transport energy in disordered organic semiconductors has been
investigated intensively. The results shows, neither Baranovskii nor
Arkhipov definition for the transport energy is valid in real
organic semiconductors system, even in the low field regime, the
transport energy lost the universality. This issue was not
adequately addressed for in earlier evaluations of the transport
energy based on the hopping rates. Concomitantly, the use of the
previously obtained expressions for transport energy in calculations
of the carrier transport parameters would lead to incorrect results
for the concentration dependencies of these parameters
\cite{baranovskii4,li4,arkhipov4,germs}. It should be mentioned that
these calculations are describing only the first release step. These
oscillating jumps of the released carriers may jump back to their
initial state do not contribute effectively to the charge transport
\cite{arkhipov2}. Finally, we mention that the description presented
here should also be applicable to describe the temperature, electric
field and carrier concentration dependent charge transport
properties, for example mobility,
as we have done in \cite{li3}.\\

Financial support from NSFC (No. 60825403) and National 973 Program
2011CB808404 is acknowledged.

\end{document}